\documentclass[12pt]{article}
\usepackage{graphicx}

\def\napoli{Faculty of Mathematics and Natural Sciences, University of Rzesz\'ow,
ul.  Pigonia 1, 35-310 Rzesz\'ow \\
Institute of Nuclear Physics PAN, PL-31-342 Krak\'ow, Poland}
\def\support{\footnote{acisek@ur.edu.pl}}

\def\Title#1{\begin{center} {\Large #1 } \end{center}}
\def\Author#1{\begin{center}{ \sc #1} \end{center}}
\def\Address#1{\begin{center}{ \it #1} \end{center}}

\newcommand\pubblock{\rightline{\begin{tabular}{l} \\
           \end{tabular}}}
\newenvironment{Abstract}{\begin{quotation}  }{\end{quotation}}
\newenvironment{Presented}{\begin{quotation} \begin{center} 
             PRESENTED AT\end{center}\bigskip 
      \begin{center}\begin{large}}{\end{large}\end{center} \end{quotation}}
\def\Acknowledgements{\bigskip  \bigskip \begin{center} \begin{large}
             \bf ACKNOWLEDGEMENTS \end{large}\end{center}}

\def\beq{\begin{equation}}
\def\eeq#1{\label{#1}\end{equation}}
\def\eeqn{\end{equation}}

\def\beqa{\begin{eqnarray}}
\def\eeqa#1{\label{#1}\end{eqnarray}}
\def\eeqan{\end{eqnarray}}

\let\bar=\overbar

\def\Dslash{\not{\hbox{\kern-4pt $D$}}}
\def\dslash{\not{\hbox{\kern-2pt $\del$}}}

\def\msb{{\bar{\ssstyle M \kern -1pt S}}}

\begin{document}
\begin{titlepage}
\pubblock

\vfill
\Title{Inclusive production of vector quarkonia at the LHC}
\vfill
\Author{Anna Cisek\support and Antoni Szczurek}
\Address{\napoli}
\vfill
\begin{Abstract}
We discuss prompt production of $J/\psi$ mesons in proton-proton collisions at the LHC
within NRQCD k$t$-factorization approach using Kimber-Martin-Ryskin (KMR) and Kutak-Stasto (KS)
unintegrated gluon distribution (UGDF). We include both direct color-singlet production $gg \to J/\psi g$
as well as a feed-down from $\chi_{c} \to J/\psi \gamma$ and $\psi' \to J/\psi X$.
The corresponding matrix elements for $gg \to J/\psi$, $gg \to \psi'$ and $gg \to \chi_{c}$
include parameters of the nonrelativistic space wave functions of the quarkonia at $r=0$,
which are taken from potential models from the literature.
We calculate the ratio of the corresponding cross sections for $\chi_{c2}/\chi_{c1}$.
We compare our results with ATLAS experimental data. Differential distributions in rapidity
and transverse momentum of $J/\psi$ and $\psi'$ are calculated and compared to experimental data
of the ALICE and LHCb collaborations.
We present results for three different values of energy 2.76 TeV, 7 TeV and 13 TeV.

\end{Abstract}
\vfill
\begin{Presented}
EDS Blois 2017, Prague, \\ Czech Republic, June 26-30, 2017
\end{Presented}
\vfill
\end{titlepage}
\def\thefootnote{\fnsymbol{footnote}}
\setcounter{footnote}{0}

\section{Introduction}

There is a long-standing lack of convergence in understaning production of
$J/\psi$ quarkonia in proton-proton or proton-antiproton collisions.
Some authors think that the cross section is dominated by the color-octet contribution,
other authors believe that the color-singlet contribution dominates. 
In the present paper we wish to calculate the color-singlet contribution as well as 
possible in the NRQCD $k_t$-factorization.
In the present approach we concentrate rather on small transverse momenta of $J/\psi$ or $\psi'$ relevant for
ALICE and LHCb data \cite{Alice_2760, LHCb_7000, Alice_7000_a, Alice_7000_b, LHCb_13000}.
We expect that color-singlet contributions may dominate in this region of the phase space.
Finally $\psi'$ quarkonium also has a sizable branching fraction into $J/\psi X$ \cite{PDG}.

\section{Inclusive production of $J/\psi$ and $\psi'$ mesons in the NRQCD $k_t$-factorization approach}

\begin{figure}[htb]
\centering
\includegraphics[height=1.95in]{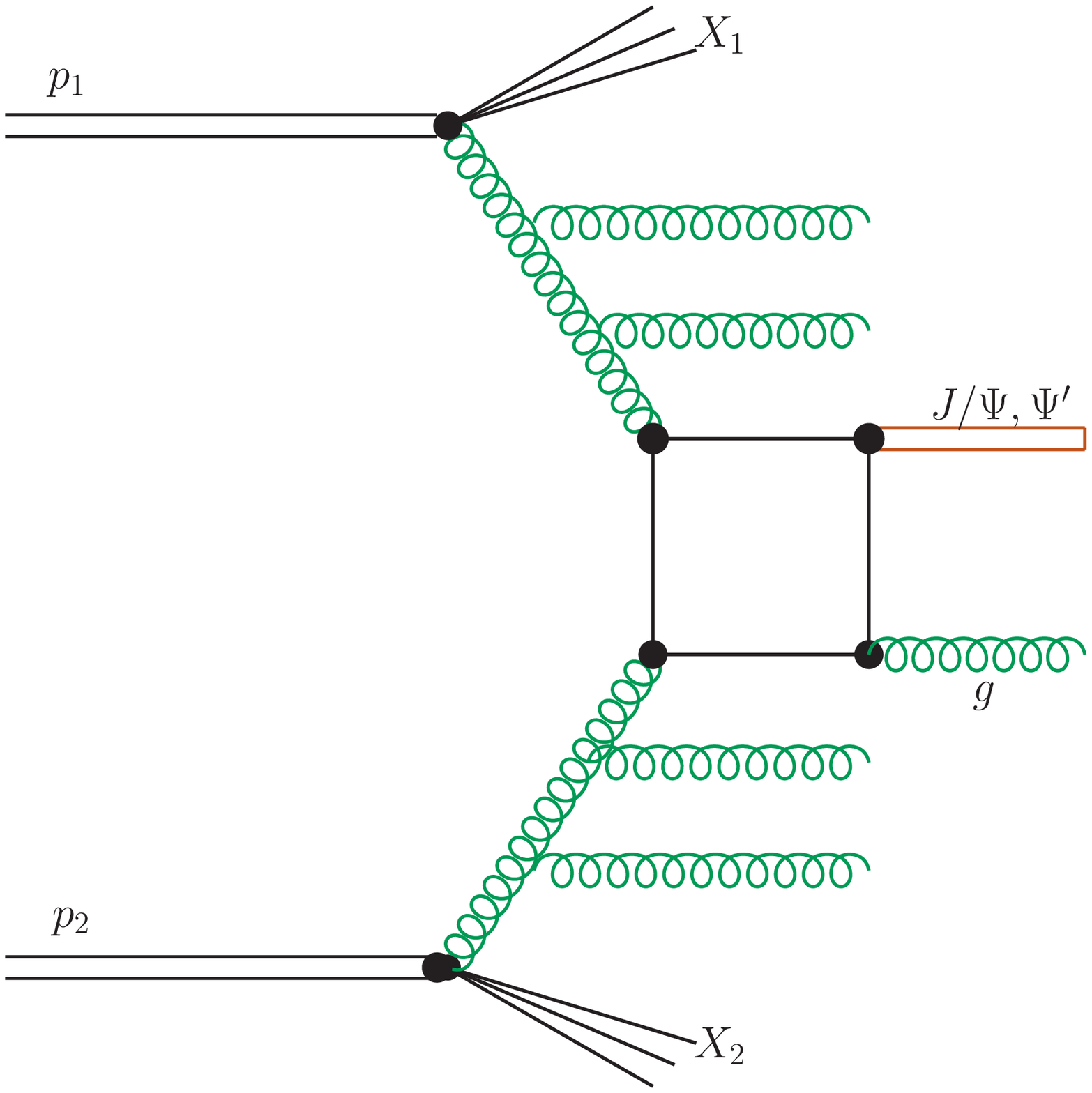}
\includegraphics[height=1.95in]{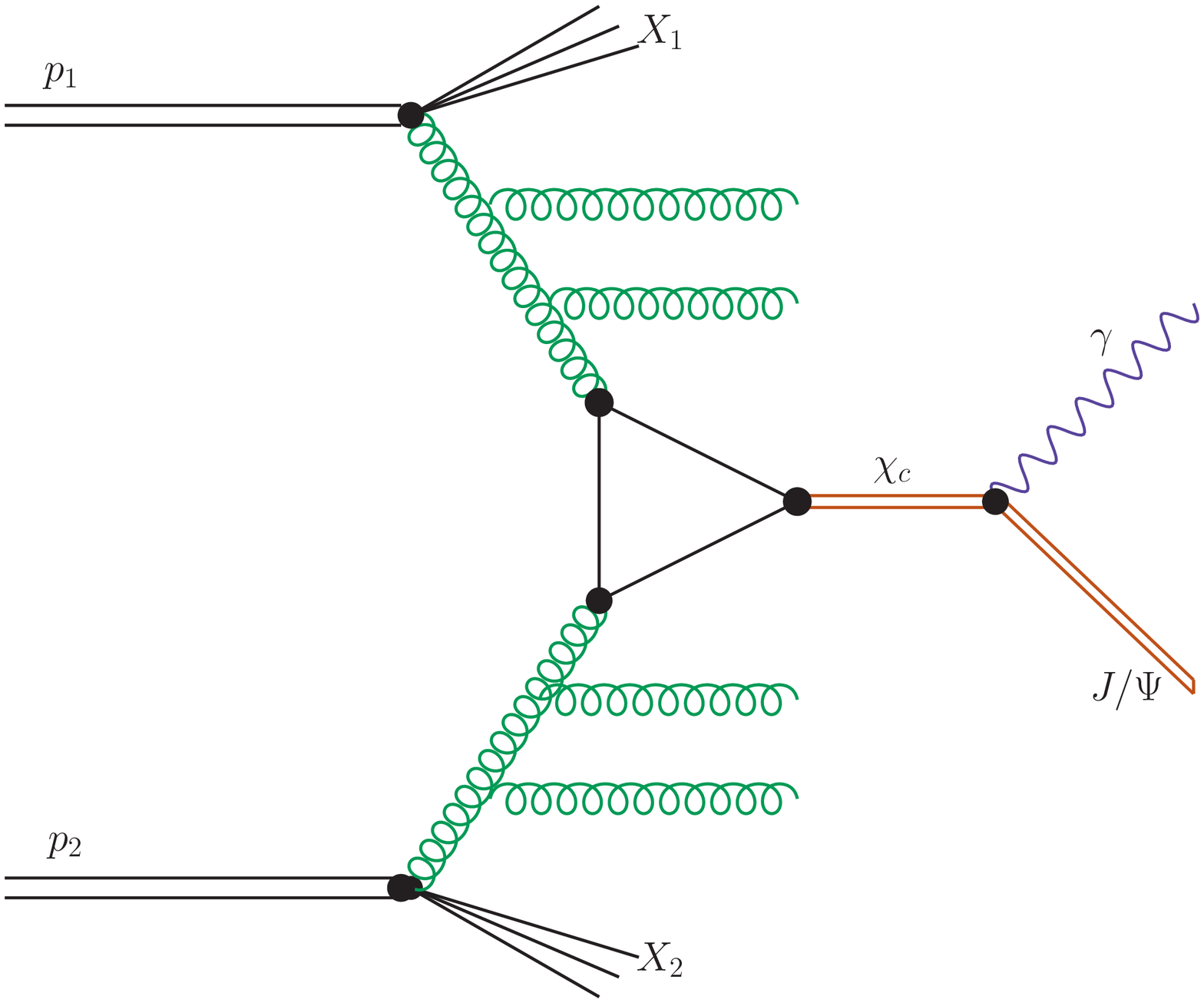}
\caption{The leading-order diagram for prompt $J/\psi$ ($\psi'$) 
meson production in the $k_t$-factorization approach.}
\label{fig_diag_1}
\end{figure}

The main color-singlet mechanism for the production of $J/\psi$ and $\psi'$ mesons is shown
in Fig.\ref{fig_diag_1} (left panel). We restrict ourselves to the gluon-gluon fusion mechanism.
In the NLO the differential cross section in the $k_{t}$-factorization
can be written as:

\begin{eqnarray}
\frac{d \sigma(p p \to J/\psi g X)}{d y_{J/\psi} d y_g d^2 p_{J/\psi,t} d^2 p_{g,t}}
&& = 
\frac{1}{16 \pi^2 {\hat s}^2} \int \frac{d^2 q_{1t}}{\pi} \frac{d^2 q_{2t}}{\pi} 
\overline{|{\cal M}_{g^{*} g^{*} \rightarrow J/\psi g}^{off-shell}|^2} 
\nonumber \\
&& \times \;\; 
\delta^2 \left( \vec{q}_{1t} + \vec{q}_{2t} - \vec{p}_{H,t} - \vec{p}_{g,t} \right)
{\cal F}_g(x_1,q_{1t}^2,\mu_{F}^{2}) {\cal F}_g(x_2,q_{2t}^2,\mu_{F}^{2}) \; .
\label{kt_fact_gg_jpsig}
\end{eqnarray}
The corresponding matrix element squared for the $g g \to J/\psi g$ is
\begin{equation}
|{\cal M}_{gg \to J/\psi g}|^2 \propto \alpha_s^3 |R(0)|^2 \; .
\label{matrix_element} 
\end{equation}
The matrix element is taken from \cite{Baranov_2002}.
In our calculation we choose the scale of the running coupling constant as:
$\alpha_s^3 \to \alpha_s(\mu_1^2) \alpha_s(\mu_2^2) \alpha_s(\mu_3^2)$ ,
where $\mu_1^2 = max(q_{1t}^2,m_t^2)$,
      $\mu_2^2 = max(q_{2t}^2,m_t^2)$ and
      $\mu_3^2 = m_t^2$,
where here $m_t$ is the $J/\psi$ transverse mass.
The factorization scale in the calculation was taken as
$\mu_F^2 = (m_t^2 + p_{t,g}^2)/2$.

In the $k_t$-factorization approach the leading-order cross section for 
the $\chi_c$ meson production can be written as:
\begin{eqnarray}
\sigma_{pp \to \chi_c} = \int d y d^2 p_t d^2 q_t \frac{1}{s x_1 x_2}
\frac{1}{m_{t,\chi_c}^2}
\overline{|{\cal M}_{g^*g^* \to \chi_c}|^2} 
{\cal F}_g(x_1,q_{1t}^2,\mu_F^2) {\cal F}_g(x_2,q_{2t}^2,\mu_F^2) / 4
\; ,
\label{useful_formula}
\end{eqnarray}
which can also be used to calculate rapidity and transverse momentum distributions 
of the $\chi_c$ mesons.
In the last equation ${\cal F}_g$ are unintegrated gluon distributions and 
$\sigma_{g g \to \chi_c}$ is $g g \to \chi_c$ (off-shell) cross section.
The situation is illustrated diagrammatically in Fig.\ref{fig_diag_1} (right panel).

The matrix element squared for the $g g \to \chi_c$ subprocess is
\begin{equation}
|{\cal M}_{gg \to \chi_c}|^2 \propto \alpha_s^2 |R'(0)|^2 \; .
\label{matrix_element} 
\end{equation}
We used the matrix element taken from Ref. \cite{KVS2006}.

For this subprocess the best choice for running coupling constant is:
$\alpha_s^2 \to \alpha_s(\mu_1^2) \alpha_s(\mu_2^2)$,
where $\mu_1^2 = max(q_{1t}^2,m_t^2)$ and
      $\mu_2^2 = max(q_{2t}^2,m_t^2)$.
Above $m_t$ is transverse mass of the $\chi_c$ meson.
The factorization scale for the $\chi_c$ meson production is fixed as $\mu_F^2 = m_t^2$.

\section{Results}

\begin{figure}[htb]
\centering
\includegraphics[height=2.25in]{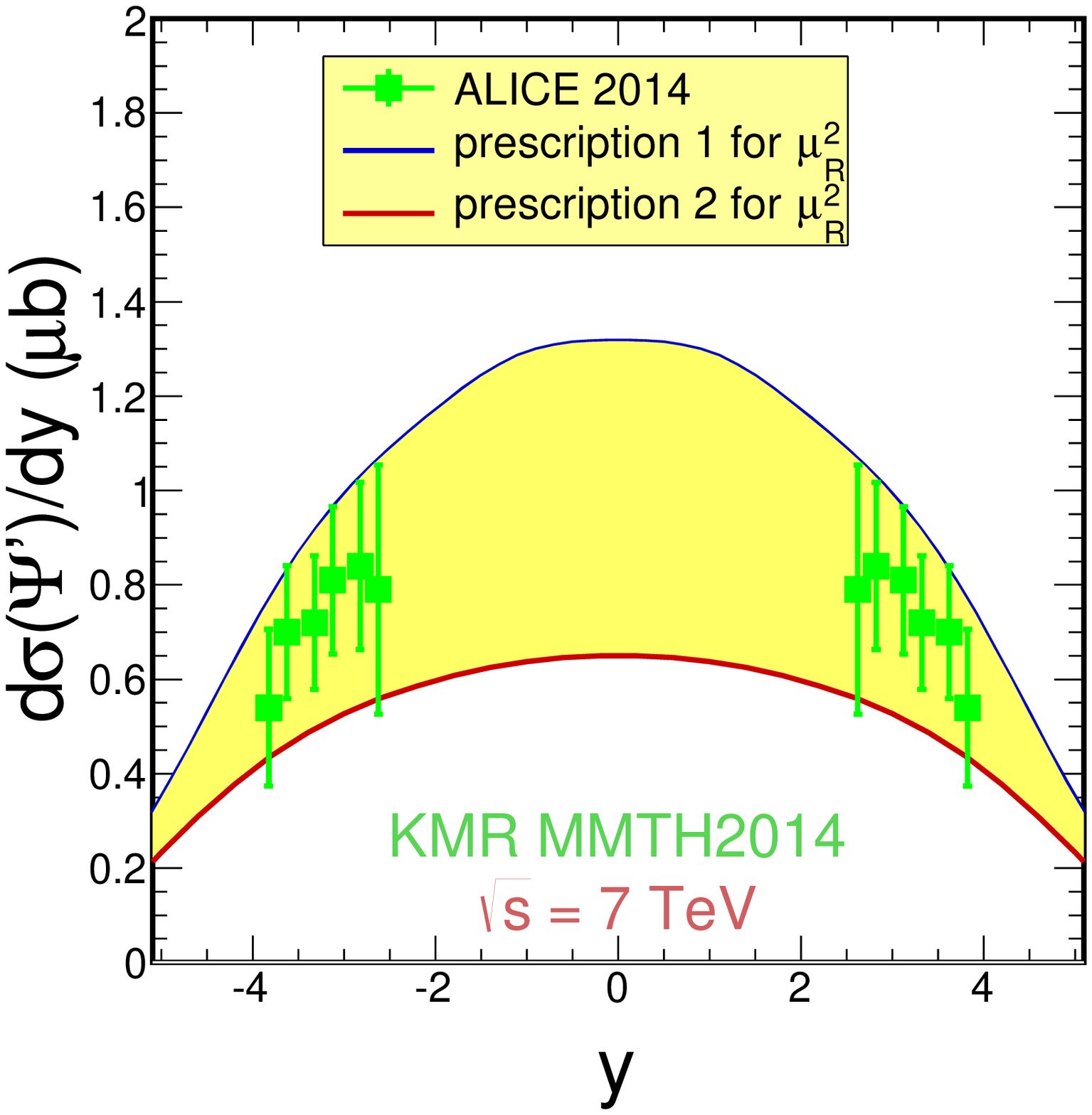}
\includegraphics[height=2.25in]{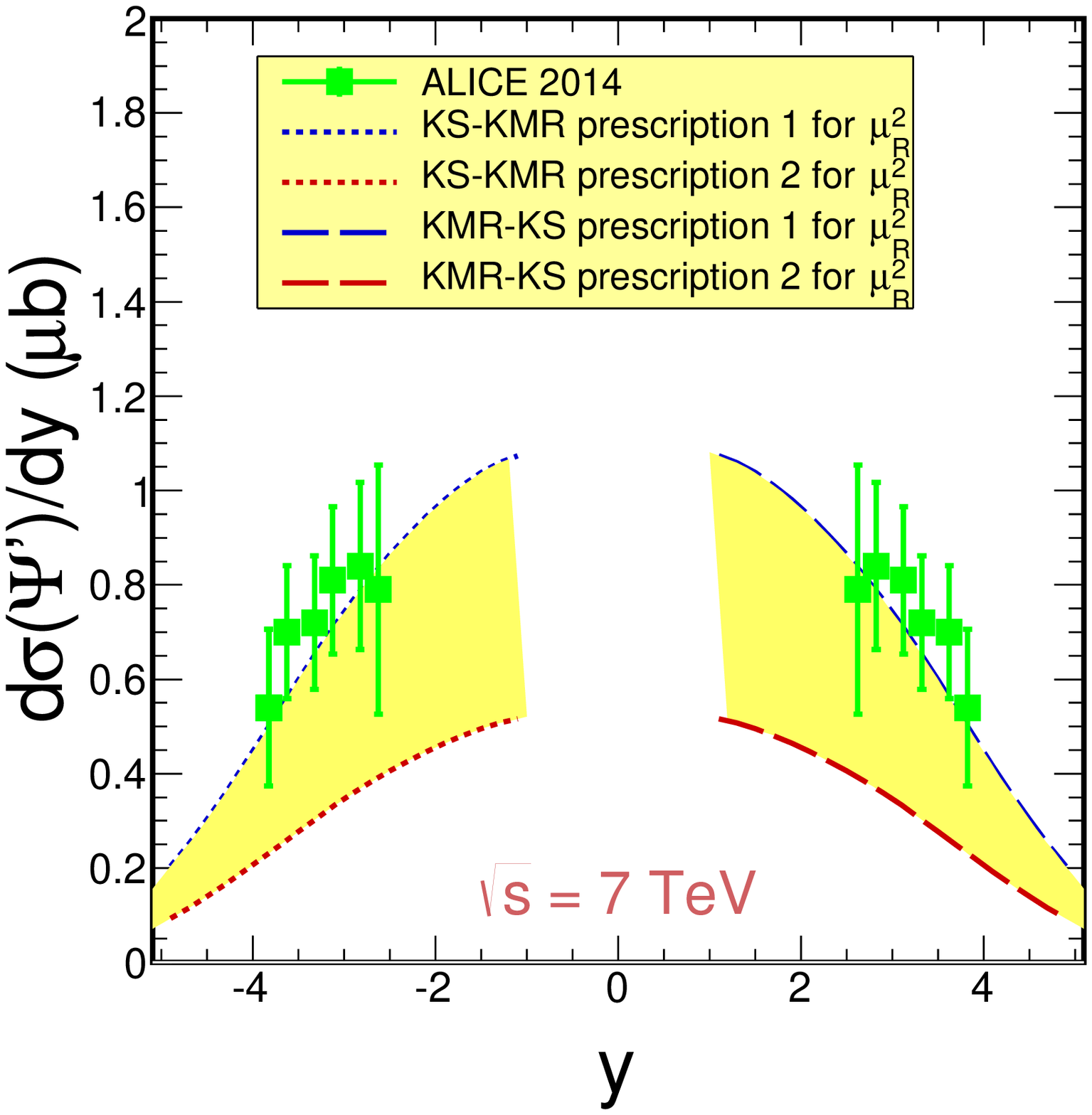}
\caption{Rapidity distribution of $\psi'$ meson with the KMR (left plots) and mixed UGDFs
(KS and KMR, right plots). The ALICE data \cite{Alice_7000_b} are shown for comparison.}
\label{fig_dsig_dy_psi2S}
\end{figure}

In Fig.\ref{fig_dsig_dy_psi2S} we show differential cross section in
rapidity for $\psi'$ production at 7 TeV.
Our results are compared with ALICE experimental data \cite{Alice_7000_b}. In the left panel we present results
for the Kimber-Martin-Ryskin (KMR) UGDF and in the right panel for mixed Kimber-Martin-Ryskin (KMR) and Kutak-Stasto (KS) UGDFs.
Because the KMR UGDF alone overshoots experimental data for rapidity distribution the best solution is to take the KMR
distribution for large x and KS for small x. For $\psi'$ meson we have to include only the direct diagram
so it's easy to compare our result with experimental data.

\begin{figure}[htb]
\centering
\includegraphics[height=1.75in]{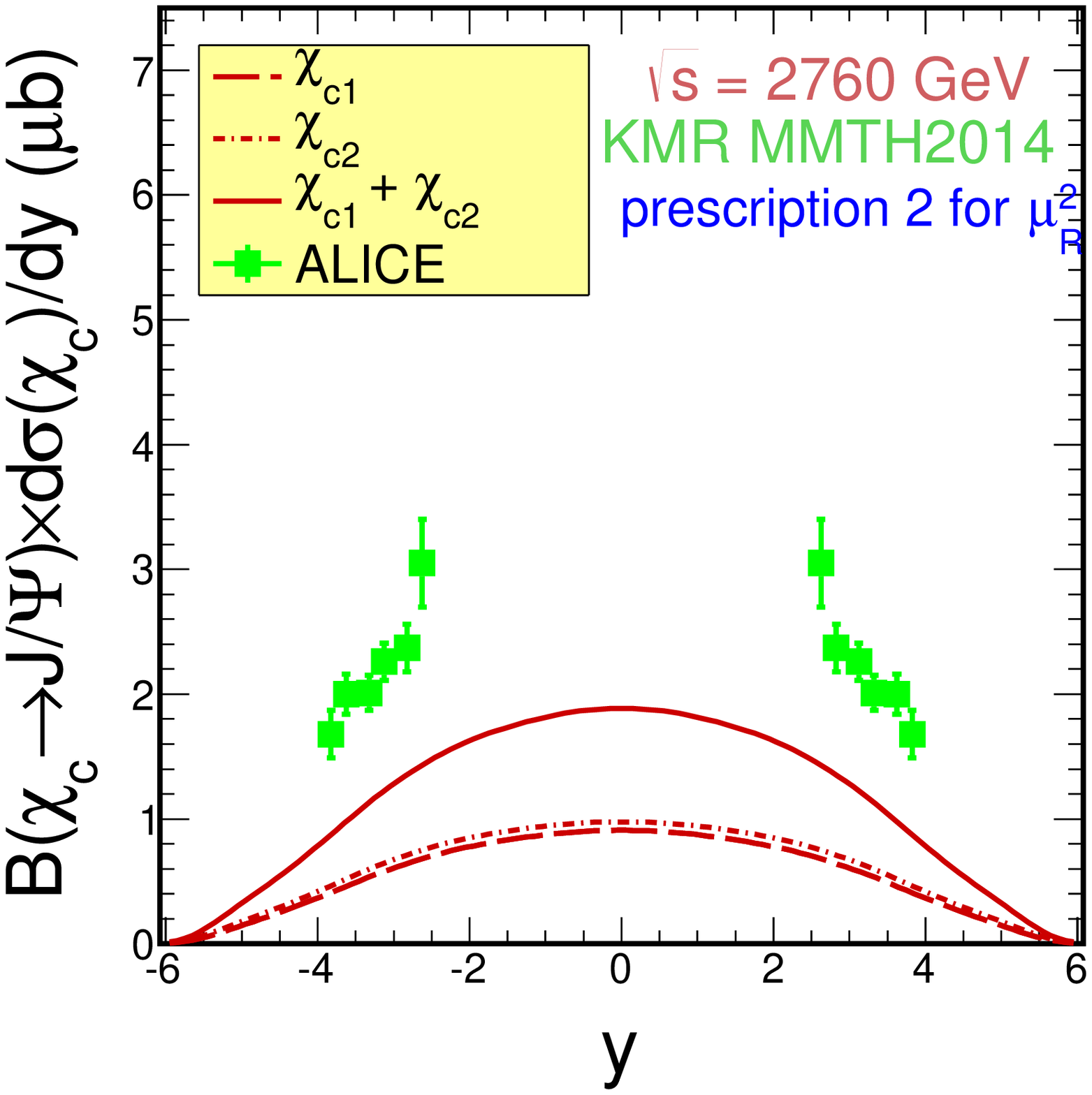}
\includegraphics[height=1.75in]{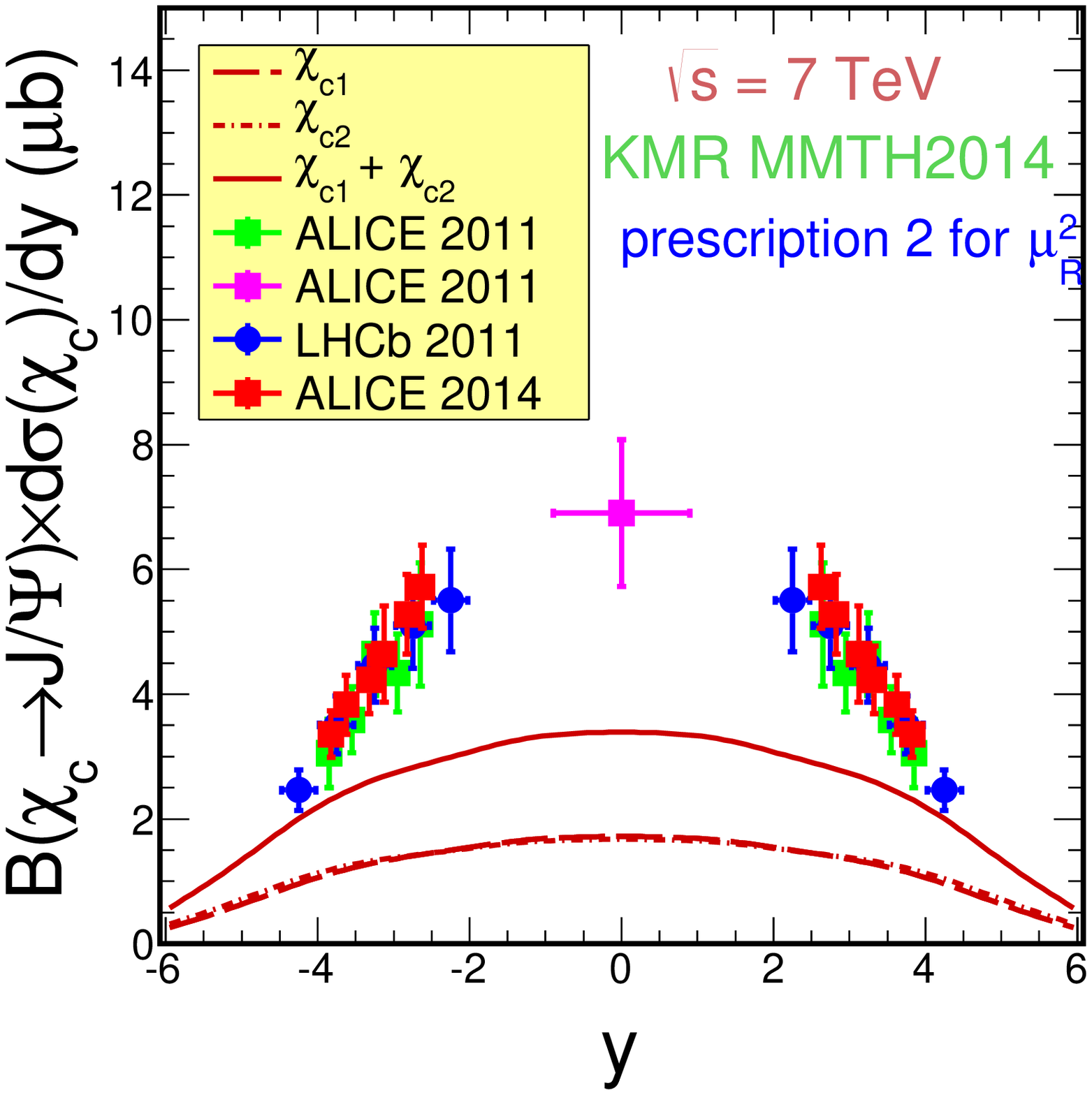}
\includegraphics[height=1.75in]{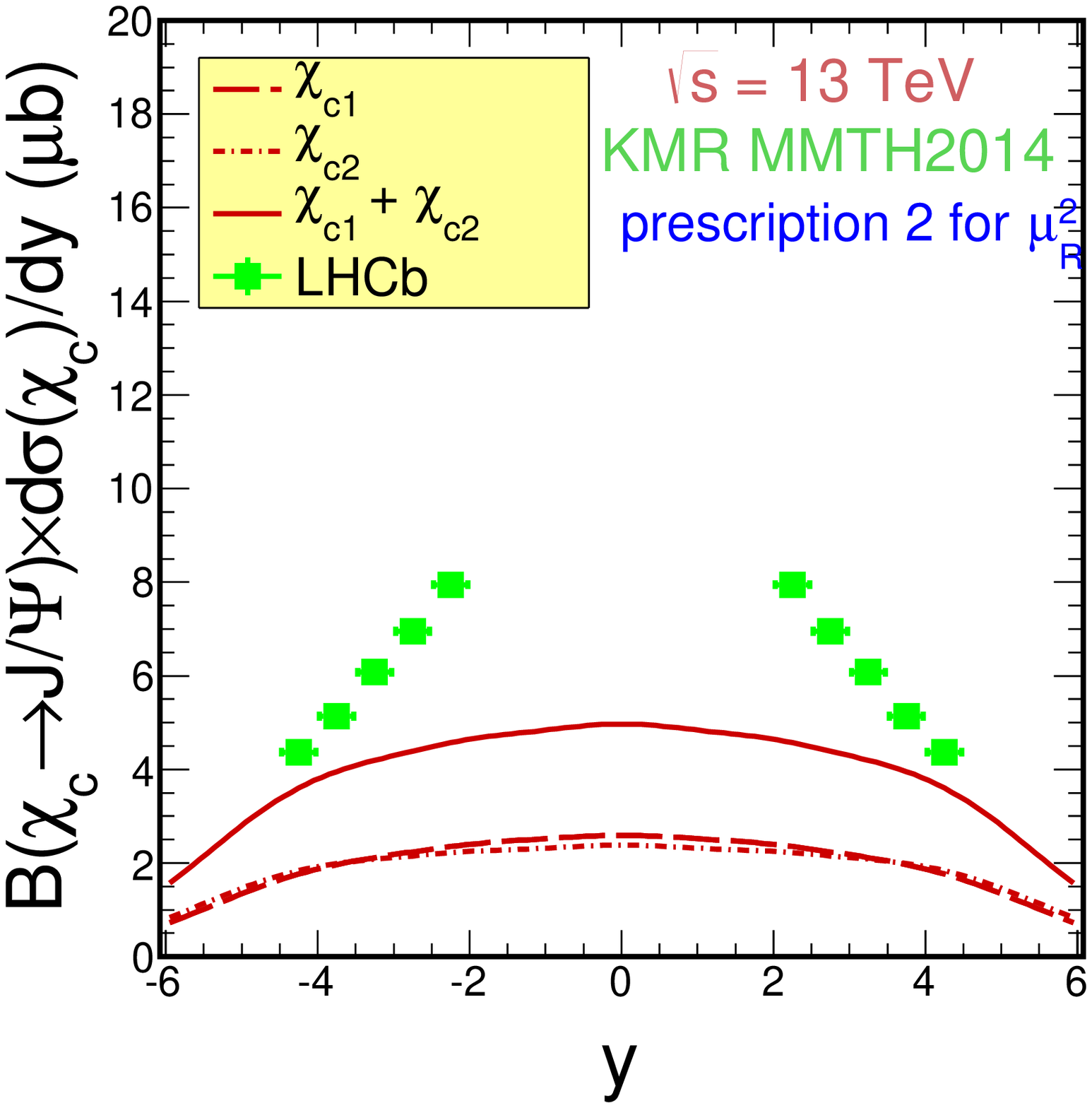}
\caption{Rapidity distribution $J/\psi$  meson from $\chi_{c}$ deacy with KMR UGDF.
The ALICE and LHCb data \cite{Alice_2760, LHCb_7000, Alice_7000_a, Alice_7000_b, LHCb_13000}
are shown for comparison.}
\label{fig_dsig_dy_jpsi_chic}
\end{figure}

In Fig.\ref{fig_dsig_dy_jpsi_chic} we present results for three different values of energy:
W = 2.76 TeV (left), W = 7 TeV (middle) and W = 13 TeV (right).
The presented results are calculated here with the KMR UGDF.
The dotted lines are for $\chi_{c1}$ meson contribution, the dot-dashed lines are for $\chi_{c2}$
meson contributions and the solid lines are a sum of these two components. 

\begin{figure}[htb]
\centering
\includegraphics[height=1.75in]{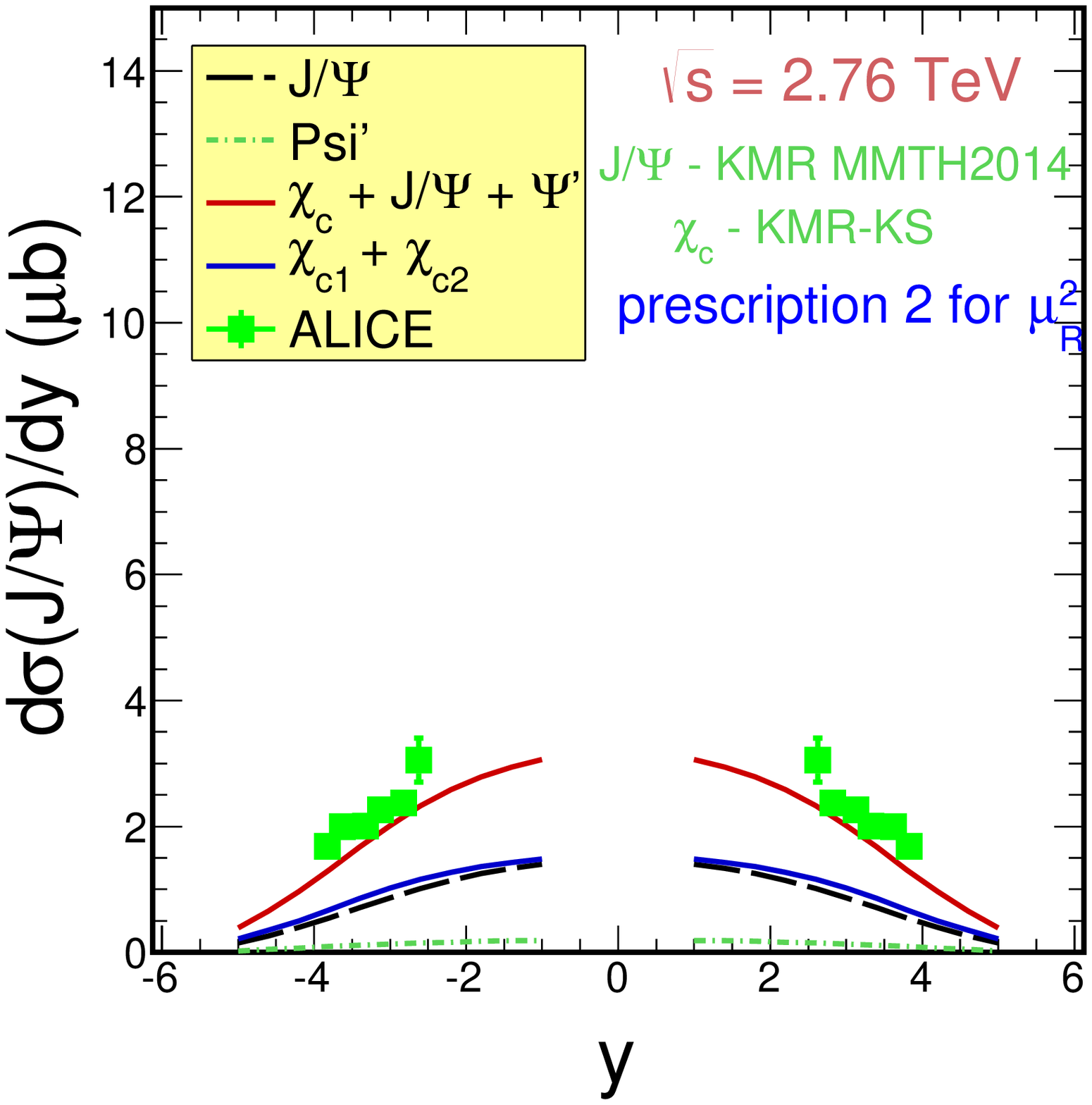}
\includegraphics[height=1.75in]{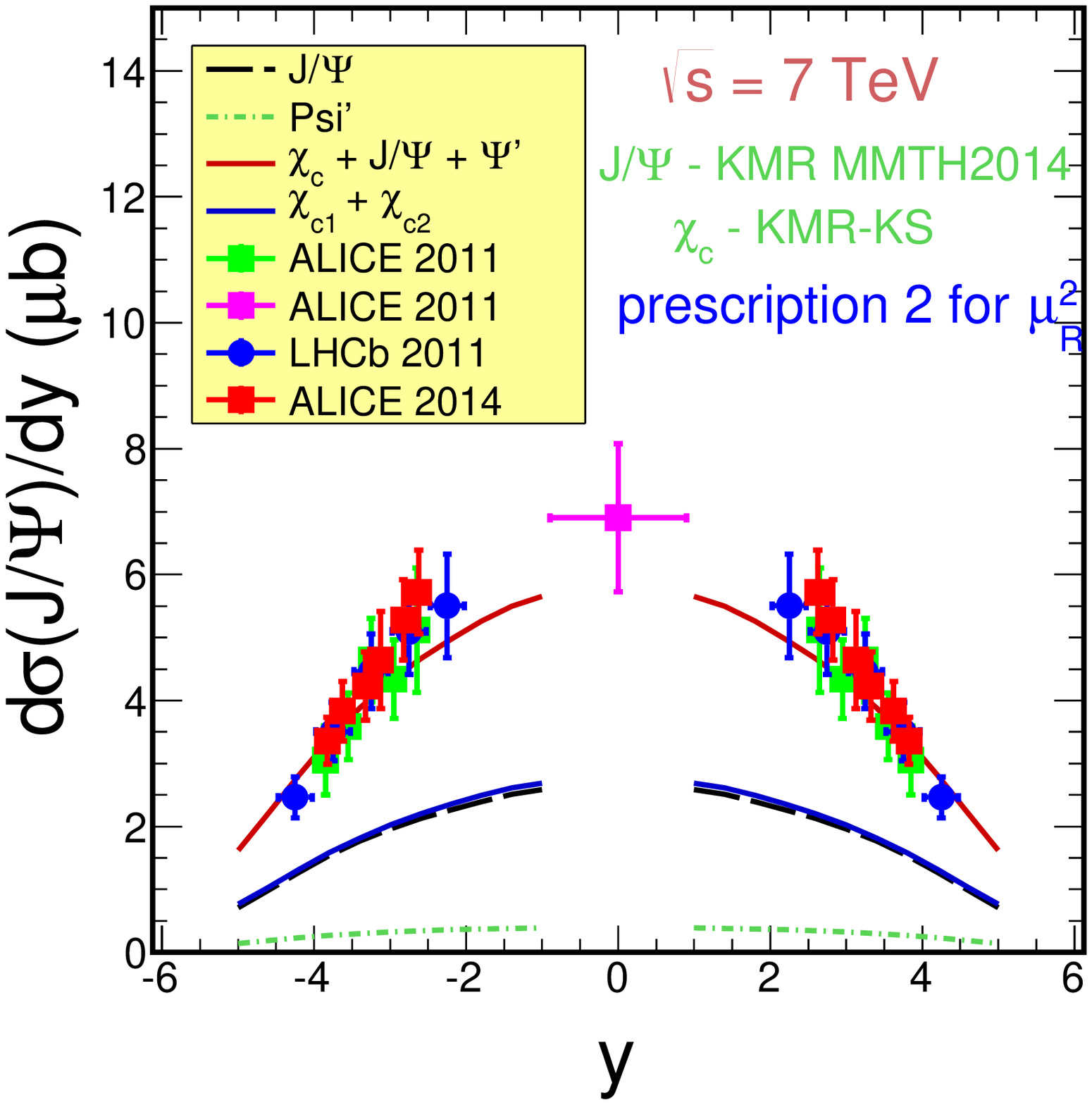}
\includegraphics[height=1.75in]{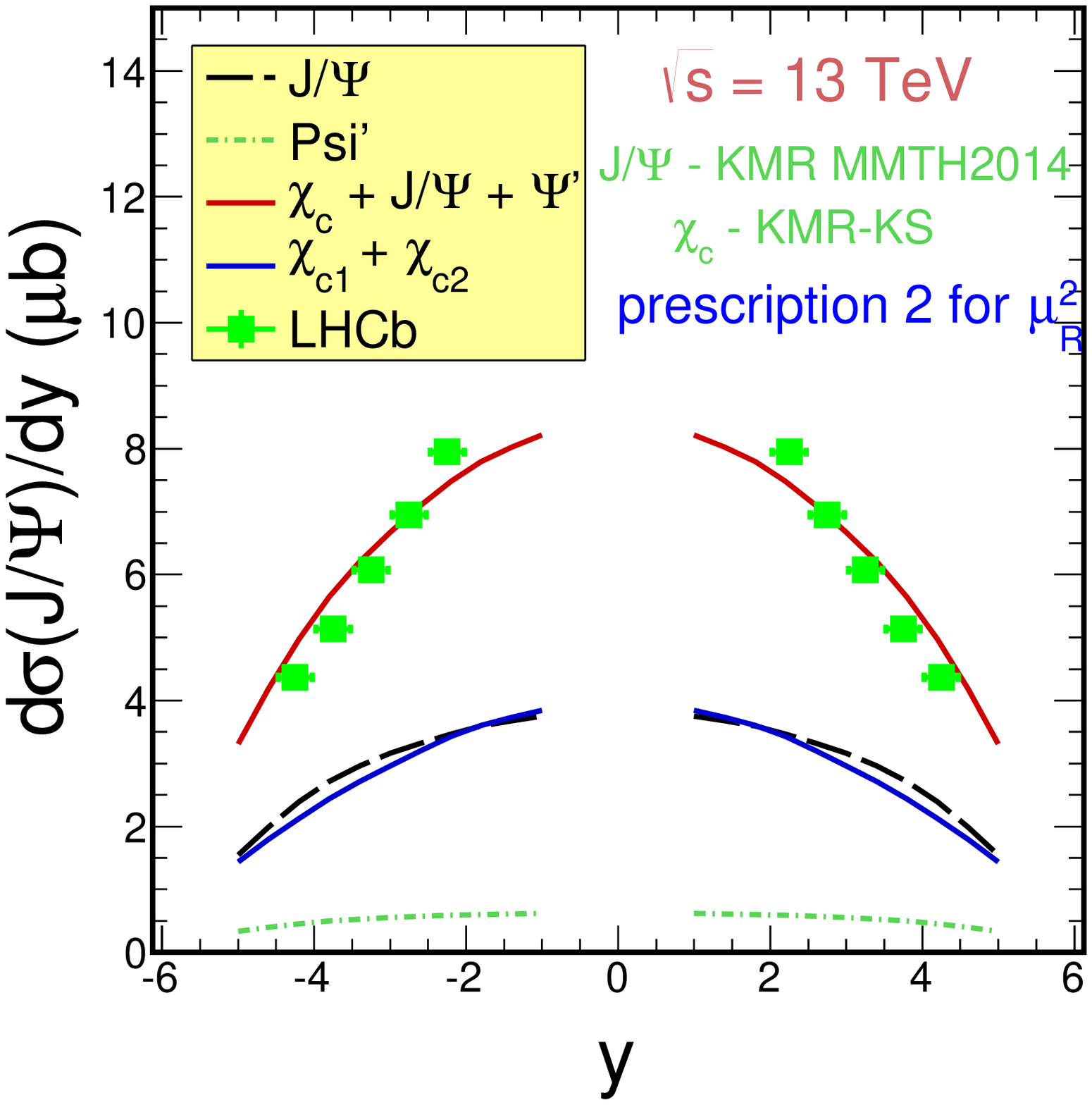}
\caption{Rapidity distribution of $J/\psi$ meson with the KMR and mixed UGDFs (KS and KMR).
The ALICE and LHCb data \cite{Alice_2760, LHCb_7000, Alice_7000_a, Alice_7000_b, LHCb_13000}
are shown for comparison.}
\label{fig_dsig_dy_jpsi_all}
\end{figure}

In Fig.\ref{fig_dsig_dy_jpsi_all} we present results for three different values of energy:
W = 2.76 TeV (left), W = 7 TeV (middle) and W = 13 TeV (right).
The presented results are calculated here with the KMR UGDF for $\psi'$ and $J/\psi$
meson direct contribution and mixed UGDF (KMR and KS) for $\chi_{c}$ meson.
The green dot-dashed lines are for $\psi'$ meson contribution.
The black dashed lines are for $J/\psi$ direct contribution.
The solid blue lines are for $\chi_{c1}$ and $\chi_{c2}$ mesons and the solid red lines are for all components.

\section{Conclusion}

We have calculated the color-singlet contribution in the NRQCD $k_t$-factorization and compared our results
with ALICE and LHCb data. Our results in rapidity are almost consistent
or even exceed  experimental data.
Cross section strongly depends on UGDF and we think the best solution is to use mixed UGDFs (KMR-KS).
Data at 13 TeV may require saturation effects in the small-x gluon.
In our approach only a small room is left for color-octet contribution.

\Acknowledgements
We would like to thank Wolfgang Sch\"afer for useful advice and discussion.
This study was partially supported by the Polish National Science Center grant 
DEC-2014/15/B/ST2/02528 and by the Center for Innovation and Transfer of
Natural Sciences and Engineering Knowledge in Rzesz\'ow.

\end{document}